\documentclass[fleqn,usenatbib]{mnras}

\usepackage{newtxtext,newtxmath}

\usepackage[T1]{fontenc}
\usepackage{pdflscape}	
\DeclareRobustCommand{\VAN}[3]{#2}
\let\VANthebibliography\thebibliography
\def\thebibliography{\DeclareRobustCommand{\VAN}[3]{##3}\VANthebibliography}

\usepackage{graphicx}	
\usepackage{amsmath}	

\title[Proton heating estimates in near-Earth CMEs]{Proton heating estimates from near-Earth observations of coronal mass ejections in solar cycle 24}

\author[D. Bhattacharjee et al.]{
Debesh Bhattacharjee,$^{1}$\thanks{E-mail: debesh.bhattacharjee@glasgow.ac.uk}
Prasad Subramanian,$^{2}$
Saikat Majumder$^{3}$
and Wageesh Mishra$^{4}$
\\
$^{1}$School of Physics and Astronomy,
University of Glasgow,
Glasgow, G12 8QQ, UK\\
$^{2}$Department of Physics,
Indian Institute of Science Education and Research Pune,
Dr. Homi Bhabha Road, Pashan, Pune-411008, India\\
$^{3}$Digantara Research and Technologies Pvt. Ltd, Hebbal Kempapura, Bengaluru, Karnataka-560024, India\\
$^{4}$Indian Institute of Astrophysics, 100 Feet Rd, Santhosapuram, 2nd Block, Koramangala, Bengaluru, Karnataka 560034, India
}

\date{Accepted XXX. Received YYY; in original form ZZZ}

\pubyear{\the\year{}}

\begin{document}
\label{firstpage}
\pagerange{\pageref{firstpage}--\pageref{lastpage}}
\maketitle

\begin{abstract}
As solar coronal mass ejections (CMEs) propagate through the heliosphere, they expend energy in heating protons to compensate for the cooling that occurs due to expansion. CME propagation models usually treat energy dissipation implicitly via a polytropic index ($\delta$). 
Here we calculate the power dissipation implied by a given $\delta$ and compare it with the power available in the turbulent velocity fluctuations. We make this comparison using near-Earth {\em in-situ} observations of 27 of the most geoeffective CMEs ($D_{\rm st} < -75$ nT) in solar cycle 24. For $\delta = 5/3$, the power in the turbulent velocity fluctuations is $\approx 54$\% smaller than what would be required to maintain the proton temperature at the observed values. If the power in the turbulent cascade is assumed to be fully expended in local proton heating, the most probable value for $\delta$ is 1.35. Our results contribute to a better understanding of CME energetics, and thereby to improved CME propagation models and estimates of Earth arrival times.
\end{abstract}

\begin{keywords}
Sun: coronal mass ejections (CMEs) -- Physical Data and Processes: turbulence -- methods: data analysis
\end{keywords}



\section{Introduction} \label{sec:intro}
Solar coronal mass ejections (CMEs) aimed at the Earth are some of the primary drivers of space weather disturbances. It is therefore no surprise that predictions of CME arrival times and speeds at the Earth constitute one of the most active fields of research in heliophysics \citep{2001Gopal,2018Riley}. Models of CME propagation through the heliosphere use observations of near-Sun CME launch times and speeds to predict when they will arrive at the Earth. The energy contained in the CME magnetic fields is usually considered to be the primary reservoir \citep{2010Mandrini}, which is expended in i) driving the CME through the heliosphere, overcoming losses due to aerodynamic drag with the ambient solar wind and ii) in compensating for the cooling of the CME plasma as it expands outwards. Unfortunately, the relative importance of these processes is not very well known. Consequently, basic model parameters need to be tuned in order to obtain approximate agreement with CME Earth arrival times. In this paper, we focus on item ii) - local heating of the CME plasma (technically, only the protons) as it propagates and expands outward. It is well-known that the CME plasma is subjected to local heating as it propagates outwards. This is borne out, for instance, by UVCS observations at $\approx 2.4 R_{\odot}$ \citep{2001Akmal,2001Ciaravella,2011Murphy,2009Lee}. 
CME propagation models typically use a one-fluid description, with no distinction between the constituents of the CME plasma, such as protons, electrons, and other heavier ions. The term ``plasma heating'' is taken to mean heating of the entire (one-fluid) plasma. Such models use a polytropic law ($P\,V^{\delta} = {\rm Constant}$) to implicitly address the CME plasma heating rate. An isothermal process ($\delta = 1$) suggests that the gas is connected with an external heat reservoir, so that there is constant heat supply, enabling it to maintain its temperature. On the other hand, for an abaibatic process ($\delta = 5/3$), the gas is thermally isolated from its surroundings and it's internal energy changes in response to the work expended in compressing or expanding the gas. Values of $\delta \approx 1$ imply that the plasma is nearly isothermal, so that the temperature is maintained at a nearly constant value owing to continuous heat transfer from the hot solar corona, without need for local heating. Conversely, $\delta \approx 5/3$ implies that the CME plasma is nearly adiabatic (i.e., there is no external heat supply), which would demand a relatively large local heating rate to compensate for the cooling of the CME as it expands. There is no consensus on the ``correct'' value of $\delta$ to use, either in one-dimensional CME propagation models \citep{1996Chen,1996KumarRust} or in 3D MHD simulations \citep{2013Lionello,2018Pomoell,2023Odstrcil,1999Ostrcil}.

Several authors use scatterplots between density and temperature from {\em in-situ} observations to calculate the polytropic index of protons. \citep{1995Totten,2020Nicolau} use this technique for the proton polytropic index in the solar wind. \cite{1993Osherovich} studied the thermodynamics inside interplanetary CMEs (ICMEs) using density-temperature scatterplots. They found that the proton polytropic index inside ICMEs is $\approx 1.2$ and electron polytropic index is $\approx 0.48$.
\cite{2022dayeh} used density-temperature scatterplots for a set of 336 events to determine the polytropic index inside ICMEs, sheath, and the pre and post-event regions. They found that inside ICMEs, the polytropic index is $\approx 0.13$ lower than the adiabatic value (5/3). \cite{1995farrugia} found that self-similar, radially expanding magnetic flux tube solutions for ICMEs require a polytropic index $< 1$, while spheromak solutions require it to be 4/3. Using ICME observations between 0.3 and 5.4 AU, \cite{2005Liu} found that the proton polytropic index in ICME protons is $\approx 1.15$ while the electron polytropic index is $\approx 0.73$. 

\cite{2009Wang}, \cite{2018Wageesh}, and \cite{2023Soumyaranjan} use the observed self-similar expansion of CMEs together with an assumption of self-similarity for the proton polytropic index to surmise that it varies from $\approx 5/3$ near the Sun to $\approx 1.2$ near the Earth. In this paper, we use the polytropic index as a free parameter and explicitly compute the predicted CME proton heating rate. We compare this prediction with the power available in the turbulent velocity fluctuation spectrum that can potentially heat the protons.

The rest of the paper is organized as follows: \S~\ref{S:CMEheating} describes the polytropic model for the temperature evolution of CME protons. We also mention the formula for the power available in the turbulent velocity spectrum in \S~\ref{S-power in turbulence}. \S~\ref{S:data} and \S~\ref{S: results} describe the data we use and the results obtained in this study, respectively. \S~\ref{S:Discussion} contains discussion and conclusions.

\section{Heating rate predicted by a polytropic process}
\label{S:CMEheating}
In this section we outline the procedure to estimate the power required to compensate for cooling of protons as a CME expands. We assume that the CME plasma follows a polytropic law 
\begin{equation}
P V^{\delta} = {\rm Constant} \, . 
\label{polytr1}
\end{equation}
where, $P$, $V$, and $\delta$ are the pressure, volume, and the ploytropic index, respectively.
We follow well-established treatments of energy conservation processes in the solar wind and obtain an expression for the local energy dissipation rate corresponding to a given value of $\delta$. The details are mentioned in the Appendix \ref{appendix}, and we quote the main result here, which concerns the evolution of proton temperature $T$:

\begin{equation}
\frac{dT(R)}{dR} = 84.32 \, (\gamma - 1)\frac{\epsilon(R)}{U(R)} - 3 (\delta - 1) \frac{T(R)}{R} \, ,
\label{nunlikeVasquez1}
\end{equation}
where $R$ is the heliocentric distance in units of solar radii (so that $R = 215$ at the Earth), $U(R)$ is the CME velocity in units of ${\rm km}\,{\rm s}^{-1}$, $\epsilon$ is the local plasma heating rate inside the CME in ${\rm J\,kg^{-1}\,s^{-1}}$, $\gamma \equiv C_{p}/C_{v}$ is the adiabatic index and $\delta$ is the polytropic index. The first term on the right hand side involving $\epsilon$ represents ``additional'' local heating due to an unspecified source. It can be due to small-scale reconnection inside the CME plasma, or due to dissipation of energy contained in turbulent fluctuations. Eq~\ref{nunlikeVasquez1} is similar to Eq 9 of \cite{2007Vasquez}, except that i) it does not presume that the CME plasma expands adiabatically; $\delta$ can take on any value, including $\gamma \equiv C_{p}/C_{v}$, and ii) the CME is assumed to expand in a self-similar manner during it's propagation, so that (as explained in the Appendix \ref{appendix}) $V \propto R^{3}$. Several observational studies show that CMEs expand self-similarly, at least beyond $\approx$ 10 $R_{\odot}$.  \citet{2014Subramanian} studied a well-observed sample of CMEs in the field of view of the Sun-Earth Connection Coronal and Heliospheric Investigation (SECCHI) coronagraphs on board the Solar Terrestrial Relationship Observatory (STEREO) satellite using the graduated cylindrical shell (GCS) model \citep{2011Thernisien} and found that CMEs propagate self-similarly as they propagate through the solar wind. A recent study of 475 CMEs observed by the STEREO satellite reveals that the ratio between the CME propagation speed and lateral expansion speed is constant, implying self-similar expansion \citep{2020Balmaceda}. 
\citet{2013Mostl} use the analytical Self-Similar Expansion Fitting (SSEF) model \citep{2012Davies} to estimate CME arrival speeds and times at the Earth. \citet{2009Wang} investigate the thermodynamics of CMEs by determining the heliocentric evolution of polytropic index, Lorentz force and gas pressure force under the assumption of self-similar expansion of CMEs.

With no additional local energy deposition inside the CME ($\epsilon = 0$), and adiabatic expansion ($\delta = \gamma$, that is, no connection to an external heat reservoir), Eq.~\ref{nunlikeVasquez1} predicts that a CME with a proton temperature of $10^{6}$ K at 1.05 $R_{\odot}$ will cool to $\approx 24$ K when it reaches the Earth. This highlights the need for additional local energy deposition (i.e., $\epsilon \ne 0$) to account for the observed temperatures of $\approx 10^{5}$ K at 1 AU, if the CME is assumed to expand adiabatically. On the other hand, if $\epsilon = 0$ (no local heating) and the plasma is isothermal ( i.e., $\delta = 1$ and it is implied that the CME plasma is connected to an external energy reservoir such as the solar corona), Eq~\ref{nunlikeVasquez1} predicts that the proton temperature remains unchanged, as expected.

As with most solar wind models, we assume that the proton temperature in the CME varies as a power law with the heliocentric distance as assumed in \citet{2009demoulintemp} and references therein.

\begin{equation}
T = T_{0} \biggl ( \frac{R}{R_{0}} \biggr )^{-\alpha} \, ,
\label{powerlaw3}
\end{equation}

where $T_{0}$ is the proton temperature at a reference heliocentric distance $R_{0}$. Using Eq~\ref{powerlaw3}, the temperature evolution equation (Eq~\ref{nunlikeVasquez1}) can be rewritten as

\begin{equation}
\epsilon (R) = [84.32\,(\gamma - 1)]^{-1} \, U(R)\, T_{0}\, R_{0}^{\alpha} \, R^{-\alpha - 1}\,\bigl [ 3 (\delta - 1) -\alpha \bigr ] \, .
\label{new1}
\end{equation}

We assume that CMEs typically start out with a proton temperature $T_{0} = 10^6$ K at $R_{0} = 1.05 R_{\odot}$. {\rm In-situ} measurements give the CME speed $U$ and proton temperature $T$ near the Earth, using which we can calculate $\alpha$ from Eq~\ref{powerlaw3}. Using Eq~\ref{new1}, we can now estimate the ``additional local'' power $\epsilon$ (${\rm J\,kg^{-1}\,s^{-1}}$) needed to heat the CME protons for a given value of the polytropic index $\delta$. We would generally expect the power requirement to be higher for adiabatic evolution ($\delta \approx 5/3$), as compared to that for isothermal evolution ($\delta \approx 1$). This is because for isothermal expansion, the CME plasma expected to be thermally connected with an external reservoir such as the solar corona. One possibility could be efficient thermal conduction along large-scale magnetic fields connecting the CME with the corona. The additional local heat input needed to maintain the CME temperature is therefore expected to be negligible. On the other hand, for adiabatic expansion, the CME is expected to be thermally isolated from it's surroundings. Therefore, there needs to be substantial additional local heat input in order to maintain the temperature. Similar treatments have been used to estimate the plasma heating rate in the solar wind \citep{2019Livadiotis}. 

\section{Power available in turbulent velocity fluctuations}
\label{S-power in turbulence}
Having estimated the power requirement for proton heating, we turn our attention to the possible source(s). Turbulence in the solar wind is extensively studied \citep{2013Bruno,2019Cranmer} and the power in turbulent fluctuations is often invoked as the source for extended heating \citep{2009ChandranHollweg,2021Smith,2023Shankarappa,2019Livadiotis}. The interiors of CMEs are also similarly turbulent \citep{2023Debesh,2024Zubair}, and turbulent fluctuations are often invoked as a source of CME heating \citep{2006Liu,2021Sorriso}. The cascade rate ($\epsilon_{t}$) represents the power per unit mass contained in the inertial range of the turbulent velocity spectrum, and this is potentially available for proton heating at small scales such as the proton inertial length or the proton gyroradius \citep{2006Liu,2021SasiApJ}. We use the well-known Kolmogorov turbulent cascade rate $\epsilon_{t} \propto \Delta U_{k}^{3} k$ (e.g., Chapter 6, \cite{2000pope}) which relates the power per unit mass ($\epsilon_t$) in the turbulent cascade to the fluid velocity fluctuations $\Delta U_{k}$ at scale $k^{-1}$. Using a dimensionless constant of proportionality $C_{0}$, the Kolmogorov cascade rate becomes
\begin{equation}
\epsilon_{t} = C_{0} \, k\, (\Delta U_{k})^{3} \, \, {\rm J\,kg^{-1}\,s^{-1}} \, , 
\label{turbpower}
\end{equation}
and we adopt $C_{0} = 0.25$ following \cite{2008HowesJGRA,2009Chandran}.
The quantity $\Delta U_{k}$ denotes the turbulent plasma velocity fluctuation at (spatial) wave-number $k$, and is evaluated as the root-mean-square (rms) velocity fluctuation inside a moving box of temporal extent $t_{\rm box}$. The wavenumber is expressed as, $k \equiv 2 \pi/(t_{\rm box} U_{\rm MO})$, where $U_{MO}$ is the running average velocity of the magnetic obstacle (MO) inside the given $t_{\rm box}$. The Magnetic Field Investigator (MFI) data we use has a time resolution of 1 minute, however, when used with the Solar Wind Experiment (SWE) data, it provides a time resolution of $\approx 92$ sec. We take two values for $t_{\rm box}$: 40 minutes and 60 minutes, respectively. These values for $t_{\rm box}$ ensure that the wavenumbers are well within the expected inertial range. 
We note that the equation for $\epsilon_{\rm t}$ used in \cite{2009Chandran} differs from ours by a factor of $\rho$ (the plasma mass density), since their definition has units of J\,cm$^{-3}$\,s$^{-1}$. The power available in the turbulent velocity spectrum ($\epsilon_{t}$, Eq~\ref{turbpower}) can be compared with the power required to heat the CME protons ($\epsilon$, Eq~\ref{new1}). Several studies \citep{2012Liang,2016Li,2024Scolini} suggest that the inertial range turbulent spectrum in ICMEs is Alfv\'enic. Although Alfv\'enic fluctuations involve fluid velocity as well as magnetic field fluctuations, we have only considered the available power in the turbulent velocity fluctuations in writing Eq~\ref{turbpower}, as \cite{2009Chandran} do, for instance. Strictly speaking, $\Delta U_{k}$ should denote the fluid velocity fluctuations perpendicular to the large-scale magnetic field, but we ignore this point for simplicity. For an Alfvenic spectrum, (assuming Kolmogorov scaling), one could also consider the available power in the turbulent magnetic field fluctuations, $\Delta B_{k}^{2}/\mu_{0} \rho$, which is on the same footing as $\Delta U_{k}^2$ (e.g., \cite{1999leamonJGR}). For Alfvenic fluctuations, $\Delta U_{k}^{2} = \Delta B_{k}^{2}/\mu_{0} \rho$. Since $\Delta U_{k}^3 = (\Delta U_{k}^2)^{3/2}$, it follows that adding the available power in turbulent magnetic fluctuations would mean replacing the quantity $C_{0}$ with $2^{3/2} C_{0}$. 

\section{Data}
\label{S:data}

For this study, we select a sample of well-observed near-Earth interplanetary coronal mass ejections (ICMEs) detected by the WIND spacecraft (\url{https://wind.nasa.gov/}) during solar cycle 24 that resulted in geomagnetic storms with $D_{\rm st} < -75$ nT. This yields a list of 27 events, which are listed in Table~\ref{T:Data}. 
The CMEs in solar cycle 24 are found to expand more than those in solar cycle 23, probably due to an overall ($\approx 40\%$) reduction of the total pressure in the heliosphere in solar cycle 24 \citep{2014Gopal}. Although the over-expansion would dilute the stored magnetic energy in these CMEs \citep{2014Gopal}, some of them still caused geomagnetic storms with $D_{\rm st}$ indices as high as -198 nT. This makes the geoeffective storms in cycle 24 an interesting dataset to study. Table~\ref{T:Data} also mentions the classification of the event based on it's magnetic field profile \citep{2016NC,2018NC}. Because of the diversity of the magnetic field configurations associated with ICMEs as observed using \emph{in-situ} spacecraft, \citet{2018NC} introduced the term `magnetic obstacles' or `MOs'. MOs are defined as plasma structures in closed magnetic loops embedded in ICMEs \citep{2018NC}. The complex magnetic structures that do not follow the definition of magnetic clouds or flux rope configuration can also be considered as MOs. The MOs associated with the ICMEs are classified into different categories depending upon how well the observed magnetic field directions fit the expectations of a static flux rope configuration \citep{2016NC, 2018NC}. Fr events indicate MOs with a single magnetic field rotation between $90^{\circ}$ and $180^{\circ}$, F+ events indicate MOs with a single magnetic field rotation greater than $180^{\circ}$, and F- events indicate MOs with a single magnetic field rotation less than $90^{\circ}$. Events labeled `complex' (Cx) have more than one magnetic field rotations while the `ejectas' (Ejs) do not have any particular rotation \citep{2019NCSoPh}. For this study, we use the 1-minute cadence data from the MFI \citep{1995mfi}, which when measured with SWE \citep{1995swe} data provide us with a time resolution of $\approx 92$s. These two instruments (MFI and SWE) are onboard the WIND spacecraft.
The WIND ICME catalog (\url{wind.nasa.gov/ICMEindex.php}) provides the event details along with the start and end times for each of the MO listed in Table~\ref{T:Data}. The plasma velocity profile as a function of time inside the MO for each of these events is provided by the WIND/SWE instrument. We use the time series of the proton temperature from the OMNI database (\url{https://omniweb.gsfc.nasa.gov/}) considering the 1-minute cadence data associated with the WIND plasma Key Parameters (KP). This is because the data from the MFI and SWE instruments (\url{https://wind.nasa.gov/mfi_swe_plot.php}) do not provide us with the time profile of proton temperature. We note that one can also derive temperature indirectly from the proton thermal speed ($v_{\rm th}$) measured by the WIND/SWE. However, this method requires assumption on whether to use $v_{\rm th}$ as the mean, rms, or most probable speed of the distribution. We compute the magnitudes of velocity fluctuation inside the ICMEs following the method adopted by \cite{2023Debesh}. The fluctuations are rms deviations around the mean inside a moving window of duration $t_{\rm box}$. The spatial wavenumber corresponding to a value of $t_{\rm box}$ is $k = 2 \pi/l$, where the length scale for a given $t_{\rm box}$ is defined as $l = U_{\rm MO}t_{\rm box}$. $U_{\rm MO}$ is the running average of plasma velocity inside the MO for a given $t_{\rm box}$ \citep{2023Debesh}. The magnitude of the corresponding velocity fluctuation in the MO is $\Delta U_{k}$. These are used in Eq~\ref{turbpower} to calculate the power in the turbulent velocity fluctuations. We use the terms ``MO" and ``CME" interchangeably, in the rest of the paper.

\section{Results}
\label{S: results}

\begin{figure}
	
	\includegraphics[width=0.5\textwidth , scale=0.9]{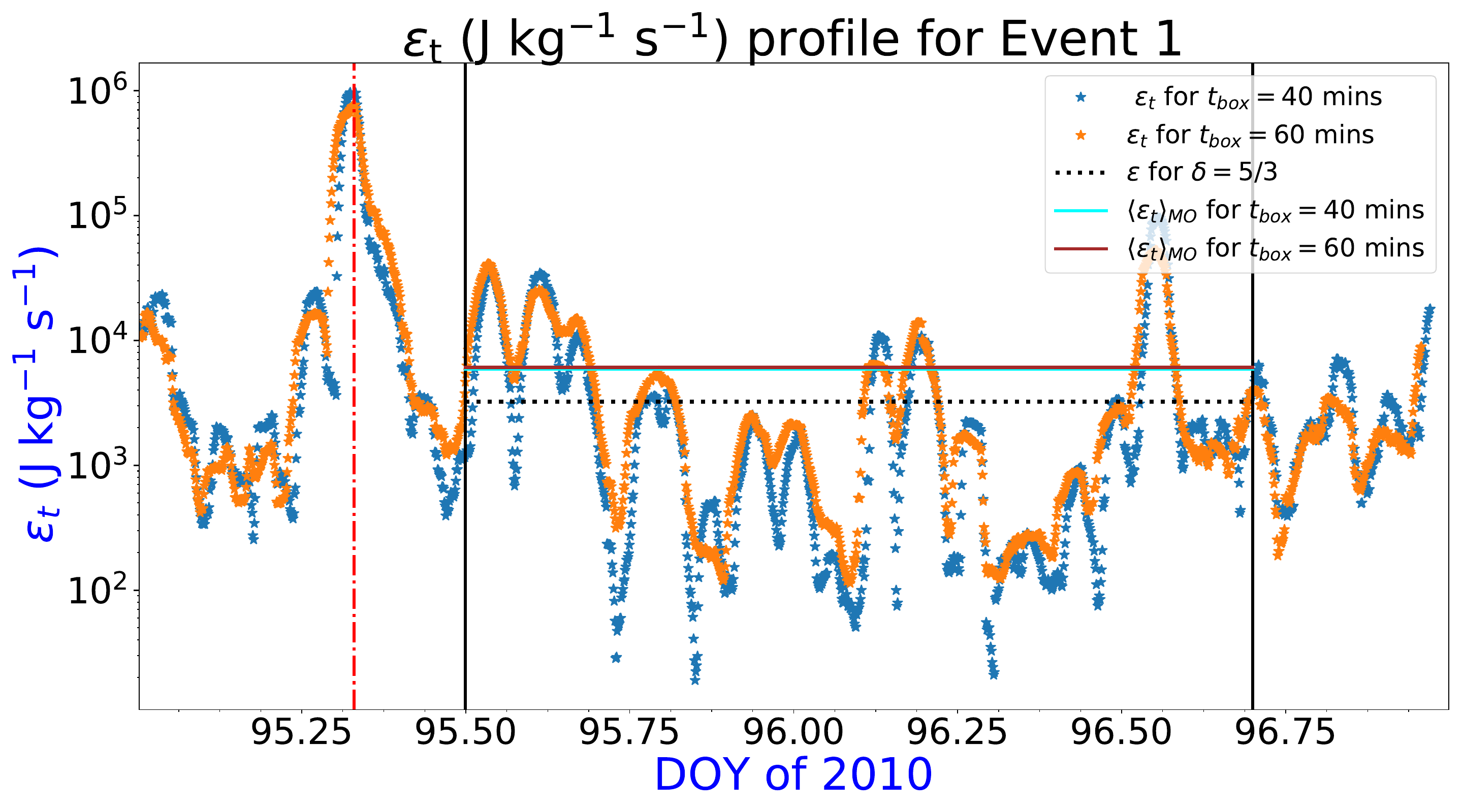}
	\includegraphics[width=0.5\textwidth , scale=0.9]{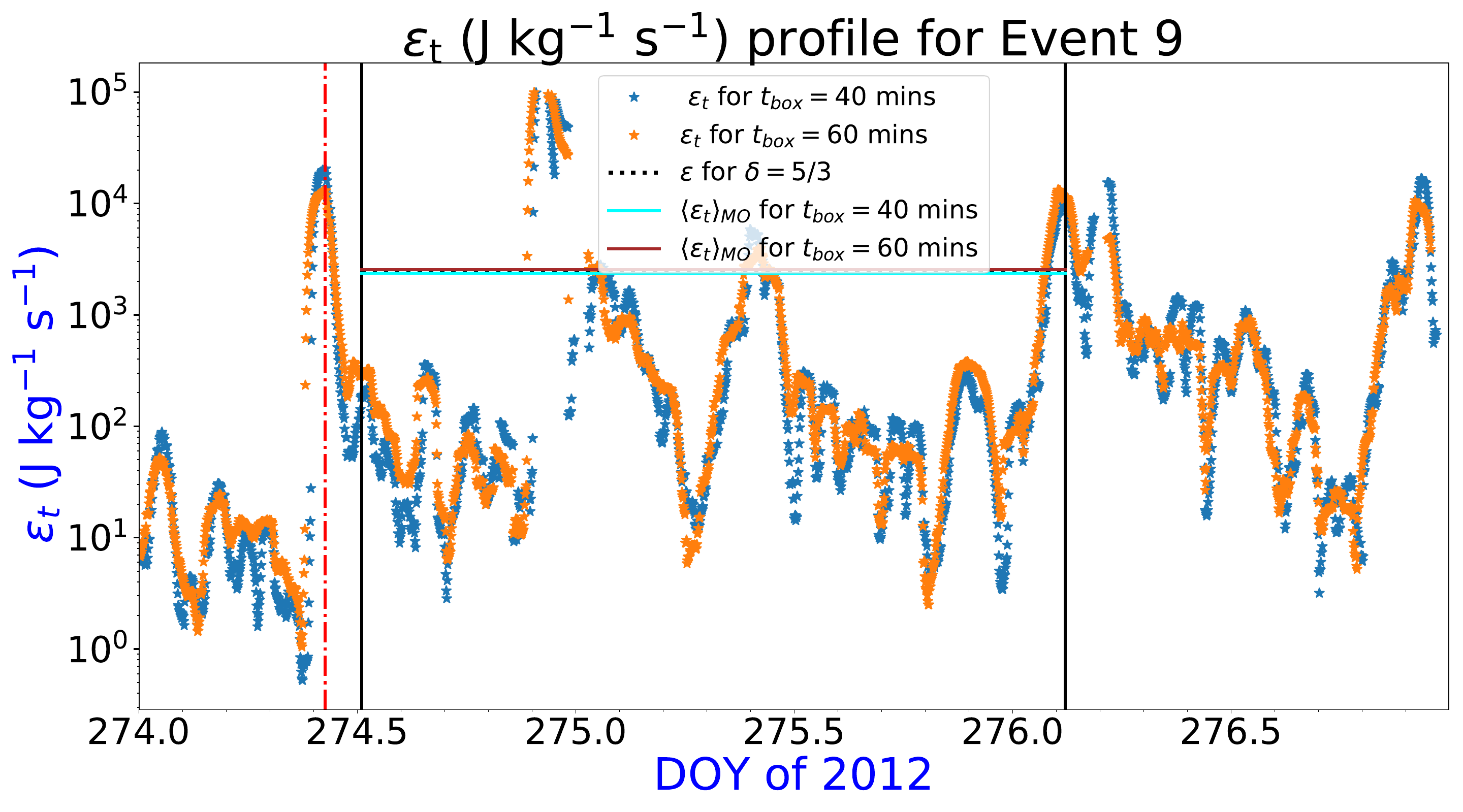}

	\caption{Time profiles of turbulent energy cascade rate ($\epsilon_{t}$, Eq~\ref{turbpower}) for event 1 (top panel) and event 9 (bottom panel) of the ICME list (Table ~\ref{T:Data}). The blue and orange scatter plots represent $\epsilon_{t}$ using $t_{\rm box} = 40$ and 60 minutes respectively. The black vertical lines mark the magnetic obstacle (MO) boundaries while the red vertical line shows the ICME start (\url{wind.nasa.gov/ICMEindex.php}). The brown horizontal line inside represents the average $\epsilon_{t}$ using $t_{\rm box} = 60$ minutes and the black dotted horizontal line represents the required power ($\epsilon$, Eq~\ref{new1}) using $\delta = 5/3$.} 
\label{Fig: events}
\end{figure}

\begin{figure}
	
	\includegraphics[width=0.5\textwidth , scale=0.9]{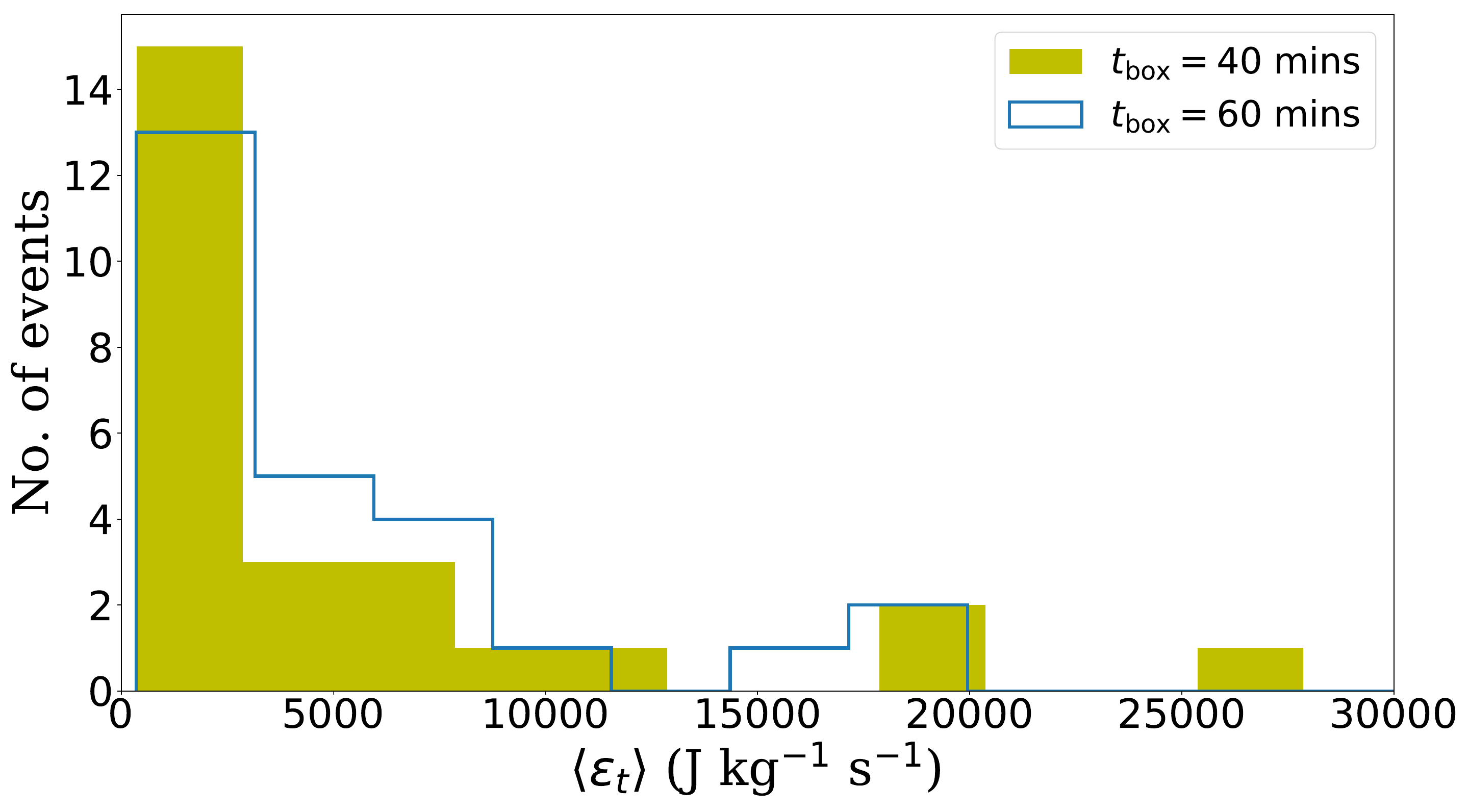}
	
	\caption{Histograms showing $\langle \epsilon_{t}\rangle$ for events listed in Table~\ref{T:Data} using $t_{\rm box} = 40$ min and $t_{\rm box} = 60$ min respectively. The mean, median, and most probable value for $t_{\rm box} = 40$ are 7646, 2674, and 1686 J kg$^{-1}$ s$^{-1}$ respectively. The mean, median, and most probable value for $t_{\rm box} = 60$ are 7750, 3486, and 1850 J kg$^{-1}$ s$^{-1}$ respectively.} 
\label{fig: epsilon_t}
\end{figure}

Figure~\ref{Fig: events} shows a time series of $\epsilon_{t}$ (Eq~\ref{turbpower}) for two representative events in our sample (Table~\ref{T:Data}) Using $t_{\rm box} = 40$ and 60 minutes \citep{2023Debesh}. We choose these two time boxes because they correspond to wavenumbers that are well within the inertial scale of the turbulent spectrum \citep{2023Debesh}, and are thus suitable to evaluate the turbulent energy cascade rate. Using $t_{\rm box} = 40$ minutes yields $\approx 55$ data points within the MO, while $t_{\rm box} = 60$ minutes yield $\approx 40$ data points. Similar plots for all the events are included in the supplementary material of this paper. Fig~\ref{fig: epsilon_t} depicts histograms of $\langle \epsilon_{t}\rangle$ for the events listed in Table~\ref{T:Data} using $t_{\rm box} = 40$ and $t_{\rm box} = 60$. The most probable value (mpv) is relevant for such skewed histograms. Since the mpv is somewhat sensitive to the histogram bin size, we use the `auto' option in matplotlib for bin size selection (following \cite{2025deep}), which chooses the smaller of the bin sizes recommended by the Sturges and Freedman-Diaconis methods. The mpv for the $t_{\rm box} = 40$ histogram is 1686 J kg$^{-1}$ s$^{-1}$ and that for the $t_{\rm box} = 60$ histogram is 1850 J kg$^{-1}$ s$^{-1}$. These numbers represent the (mpv of) power input to the protons. They may be compared with the power in the turbulent cascade that is potentially available for the dissipation on protons. Eq~\ref{turbpower} represents a simple calculation of the turbulent power, and representative results from this calculation are shown in Figure~\ref{Fig: events}. More detailed calculations of the available turbulent power in CMEs can be found, for instance, in \cite{2021Sorriso}. 

Since the gross trends for $\epsilon_{t}$ are not very sensitive to the value of $t_{\rm box}$, we use $t_{\rm box} = 60$ minutes henceforth. The brown horizontal line in Fig~\ref{Fig: events} denotes the average value of $\epsilon_{t}$ inside the MO (with $t_{\rm box} = 60$ minutes), while the black horizontal line denotes the value of $\epsilon$ (Eq~\ref{new1}) with $\delta = 5/3$ and $R = 215R_{\odot}$. For the event depicted in the upper panel of Fig~\ref{Fig: events}, $\epsilon_{t} > \epsilon$ while $\epsilon_{t} \approx \epsilon$ for the event depicted in the lower panel.

\begin{figure}
	
	\includegraphics[width=0.5\textwidth , scale=0.9]{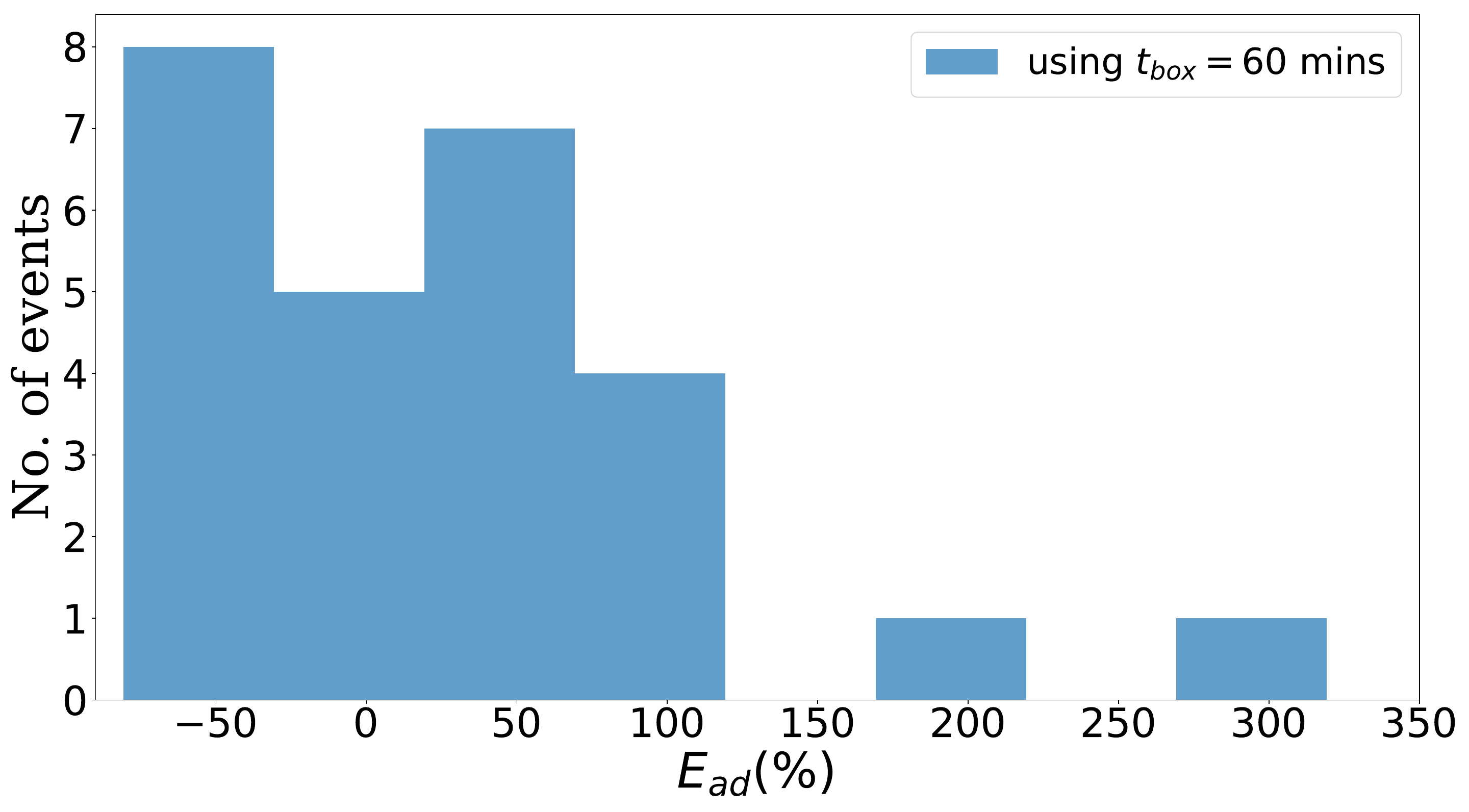}
	
	\caption{Histogram of $E_{\rm ad}$ (Eq~\ref{Ead}) for $t_{\rm box} = 60$ minutes. The mean, median, and mpv are 53.25, 25.24, and -53.8, respectively.} 
\label{figead}
\end{figure}

Next, we compare the turbulent energy cascade rate ($\epsilon_{\rm t}$) estimated inside MOs with the required proton heating rate ($\epsilon$) for $\delta = \gamma = 5/3$. This value for $\delta$ presumes adiabatic expansion and is commonly used in several studies; e.g., global 3D MHD simulations \citep{2016Wu, 2003Riley}, studies of CME-CME interaction \citep{2005LugazCMEinteraction}, studies of CME expansion \citep{2005Lugazdensity}, and many more. 
We define
\begin{equation}
E_{\rm ad} \equiv 100 \frac{\epsilon_{t} - \epsilon}{\epsilon}\, \% \, ,
\label{Ead}
\end{equation}
which is the relative (percentage) difference between the power available in the turbulent velocity fluctuations (Eq~\ref{turbpower}) and the power required for proton heating if they are assumed to remain adiabatic (Eq~\ref{new1} with $R = 215$ and $\delta = 5/3$). The histogram for $E_{\rm ad}$ for all the events in our sample using $t_{\rm box} = 60$ minutes is shown in Fig~\ref{figead}. The mpv of $E_{\rm ad}$ (which is the relevant quantity for a skewed distribution such as this one) is $-54 \%$. 
In other words, Fig~\ref{figead} suggests that the power in the turbulent velocity fluctuations is $\approx 54\%$ lower than what is required to heat the protons if they are adiabatic. Needless to say, the value of $E_{\rm ad}$ will be quite different for other assumed values of $\delta$, but $\delta = 5/3$ is a reasonable reference value considering how often it is used in the literature.

\begin{figure}
	
	\includegraphics[width=0.5\textwidth , scale=0.9]{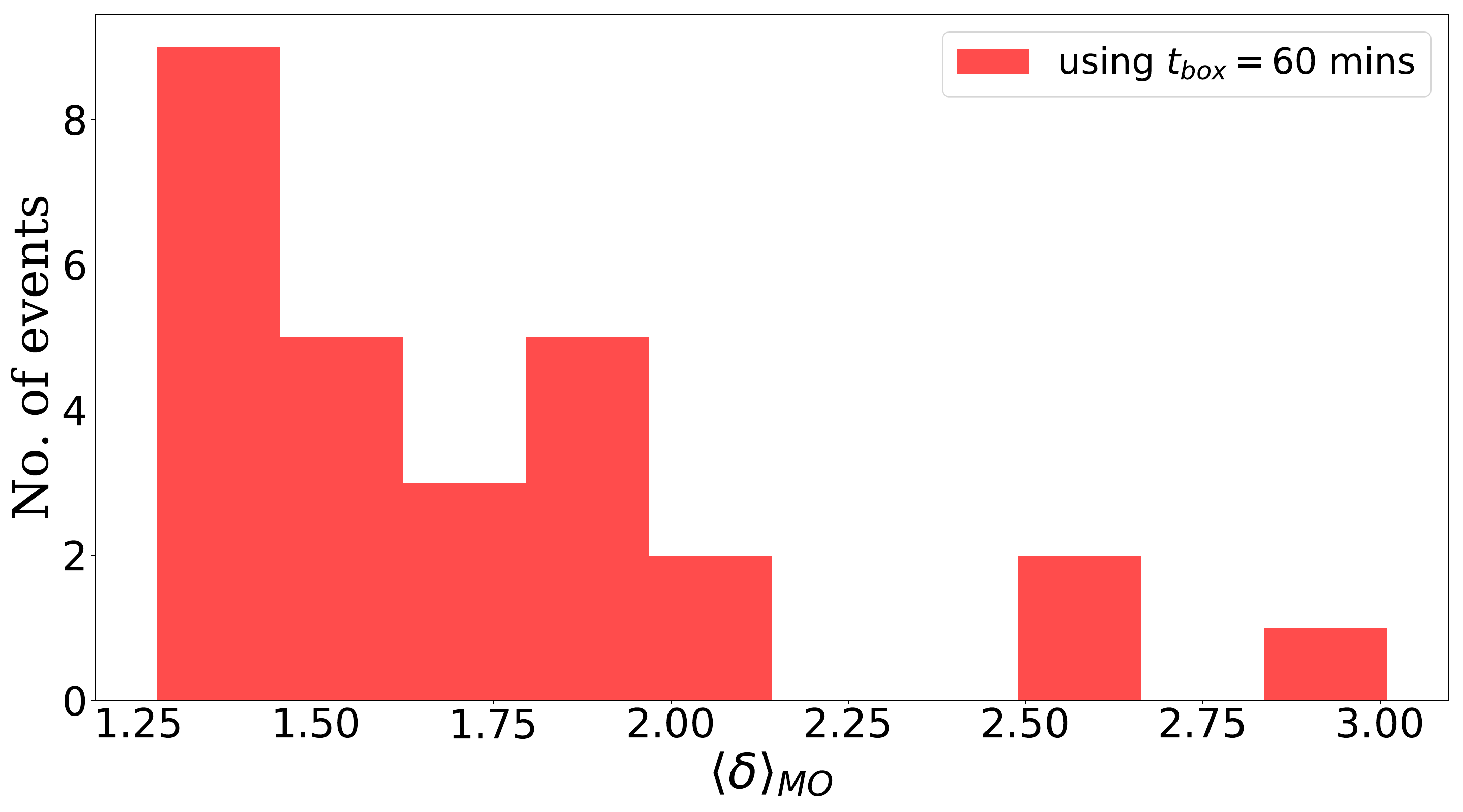}

	\caption{Histogram of estimated time-averaged polytropic index considering the turbulent energy cascade as the source of proton heating inside the near-Earth MOs ($\langle \delta \rangle_{\rm MO}$, Eq~\ref{reqdelta}) for $t_{\rm box} = 60$ minutes. The mean, median, and mpv of this histogram are 1.72, 1.59, and 1.35, respectively.} 
\label{Fig: hist}
\end{figure}

\begin{figure}
	
	\includegraphics[width=0.5\textwidth , scale=0.9]{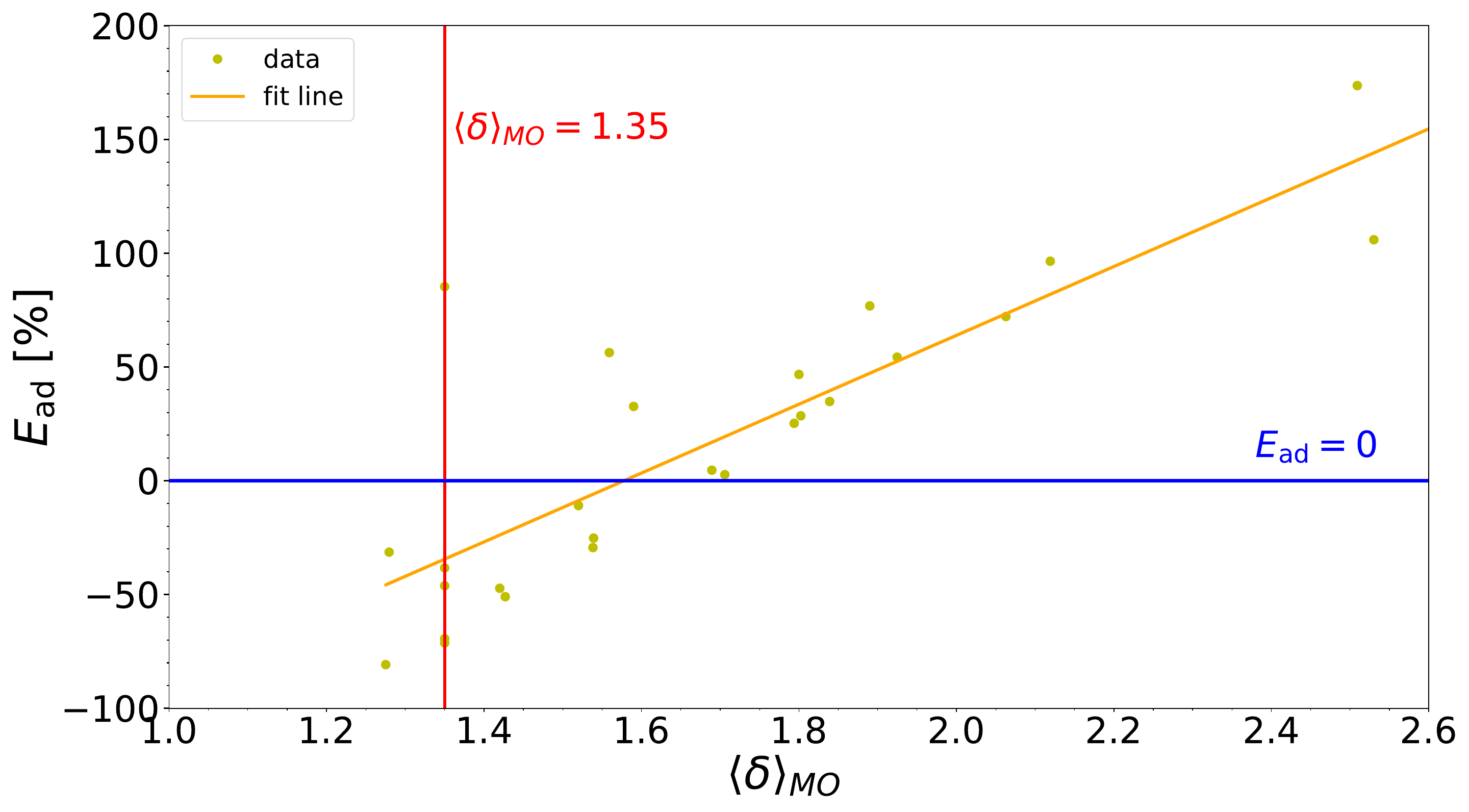}

	\caption{A scatterplot between $\langle \delta \rangle_{\rm MO}$ (Eq~\ref{reqdelta}) and $E_{\rm ad}$ (Eq~\ref{Ead}) for the events in this study. The yellow line shows the best linear fit between these two parameters. The red vertical line is at $\langle \delta \rangle_{\rm MO} = 1.35$, while the blue horizontal line denotes $E_{\rm ad}=0$.} 
\label{Fig: Ead_delta}
\end{figure}

\begin{figure}
	
	\includegraphics[width=0.5\textwidth , scale=0.9]{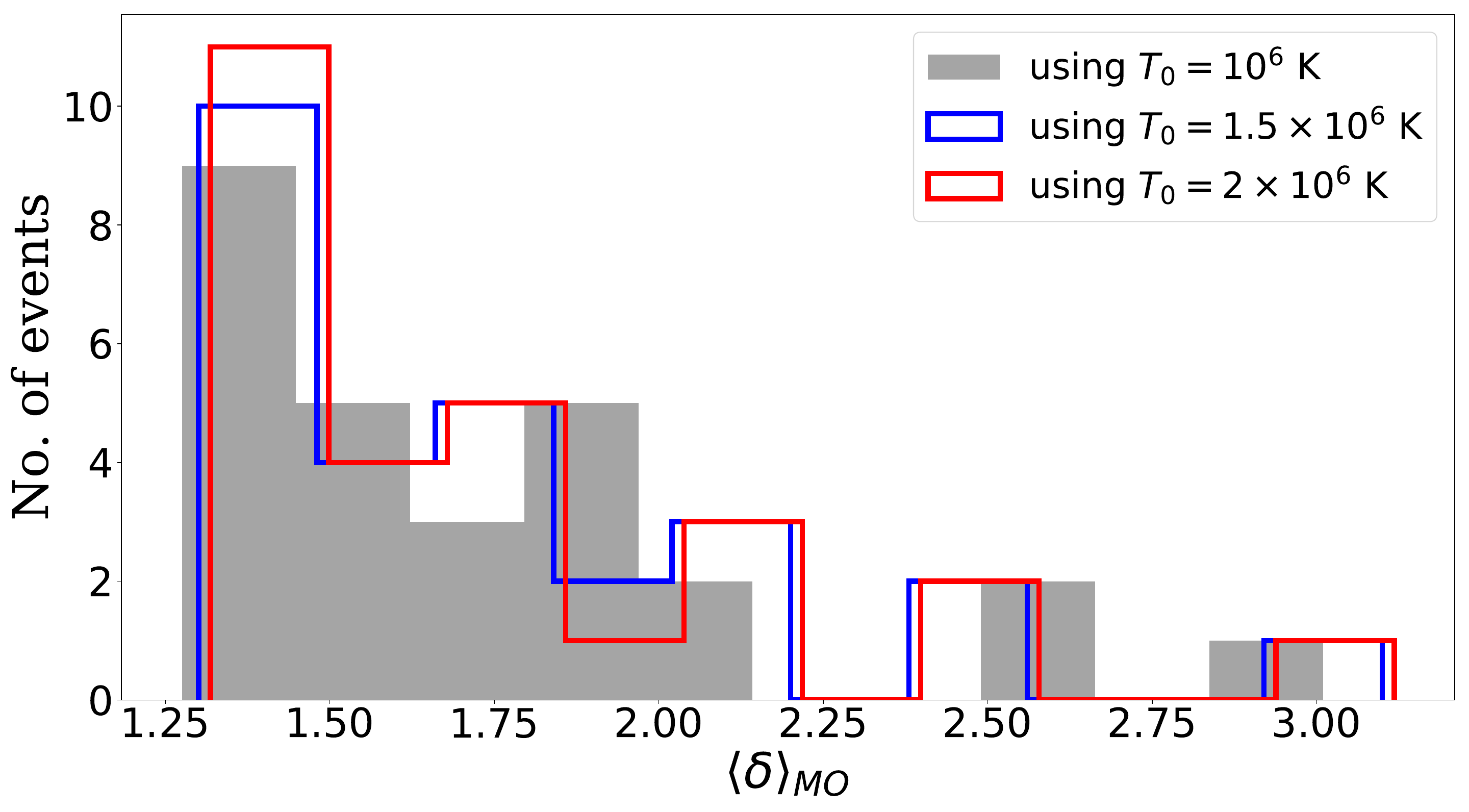}

	\caption{ Same as Fig~\ref{Fig: hist}, for different values of the initial temperature $T_{0}$. For $T_0 = 10^6$K: the mean, median, and mpvs are 1.72, 1.59, and 1.35, respectively; for $T_0 = 1.5 \times 10^6$K: the mean, median, and mpvs are 1.74, 1.60, and 1.37, respectively; and for $T_0 = 2 \times 10^6$K: the mean, median, and mpvs are 1.75, 1.61, and 1.39, respectively.} 
\label{Fig: delta_t0}
\end{figure}

Nonetheless, it is not (apriori) clear what value should be used for the polytropic index ($\delta$) in a CME propagation model. The preceding discussion suggests that the adiabatic expansion for the protons could be too ``demanding'', in that the energy in the turbulent velocity spectrum is not enough to provide the required local heating. On the other hand, the power required for local proton heating would be low if the protons were nearly isothermal. By way of estimating the ``correct'' value for $\delta$ at a specific heliocentric distance and for a given wave-number, we equate the power in the velocity fluctuations for a certain wave-number ($\epsilon_{\rm t}$, Eq~\ref{turbpower}) to the power required for proton heating ($\epsilon$, Eq~\ref{new1}) at that specific heliocentric distance. Using the estimates of $\epsilon_{\rm t}$ and $\epsilon$ at the location of the WIND spacecraft ($\approx$ 1 AU or $R = 215 R_{\odot}$), we can therefore compute the following. 

\begin{equation}
\langle \delta \rangle_{MO} = \frac{1}{3}\,\biggl [ \alpha + 3 + 84.32 (\gamma - 1) (U(R) \,R_{0}^{\alpha}\,T_{0})^{-1}\,R^{\alpha + 1}\,\langle \epsilon_{t} \rangle \biggr ] \, ,
\label{reqdelta}
\end{equation}
where $\langle \epsilon_{t} \rangle$ denotes the time-averaged value $\epsilon_{t}$ inside an MO and $\langle \delta \rangle_{MO}$ the time-averaged value of the polytropic index inside the MO. 
The histogram of $\langle \delta \rangle_{MO}$ with $R = 215 R_{\odot}$ and $t_{\rm box} = 60$ minutes (Fig~\ref{Fig: hist}) shows that it ranges from 1.25 to 3, with a mpv of 1.35. We have noted that $\epsilon$ generally exceeds $\epsilon_{t}$ with $\delta = 5/3$ (Fig~\ref{figead}). It is therefore not surprising that the mpv of $\langle \delta \rangle_{MO}$ (which is obtained by equating $\epsilon$ with $\epsilon_{t}$) is lower than 5/3. 
Fig~\ref{Fig: Ead_delta} shows a scatterplot between $\langle \delta \rangle_{\rm MO}$ and $E_{\rm ad}$. 
The linear fit between $\langle \delta \rangle_{\rm MO}$ and $E_{\rm ad}$ (depicted by the yellow line) has an intercept of -39.7\% at $\langle \delta \rangle_{\rm MO} = 1.35$ (the red vertical line in Fig~\ref{Fig: Ead_delta}). In other words, at $\langle \delta \rangle_{\rm MO} = 1.35$, the power stored in the turbulent fluctuations is $\approx 40\%$ lower than the power required to heat the protons assuming adiabatic expansion.
Eq~\ref{reqdelta} shows that $\langle \delta \rangle_{MO}$ depends on the initial temperature ($T_0$) we assume. In order to check the sensitivity of $\langle \delta \rangle_{MO}$ to $T_{0}$, we compute it using three different initial temperatures: $T_0 = 10^6$K, $1.5 \times 10^6$K, and $2 \times 10^6$K, respectively and $t_{\rm box} = 60$ minutes. The results are shown in the histograms of Fig~\ref{Fig: delta_t0}. The mpv of $\langle \delta \rangle_{MO}$ increases by $\approx 3\%$ upon doubling the initial temperature (i.e., $T_0 = 2 \times 10^6$K instead of $10^6$ K). This implies that $\langle \delta \rangle_{MO}$ depends only weakly on the assumed initial temperature ($T_0$).

\begin{figure}
	
	\includegraphics[width=0.5\textwidth , scale=0.9]{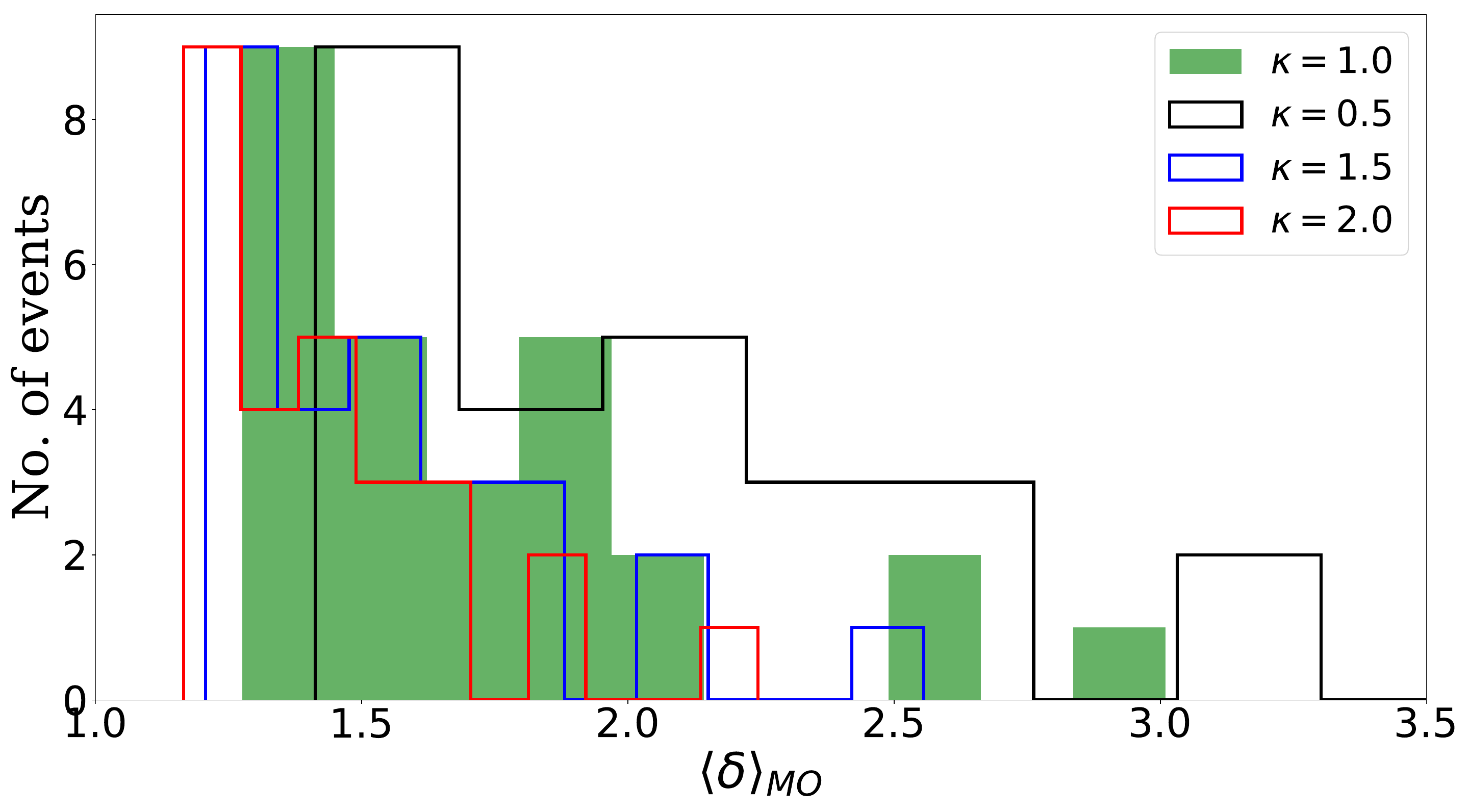}

	\caption{ Histograms of $\langle \delta \rangle_{\rm MO}$ for different values of $\kappa$. For, $\kappa=0.5$, the mean, median, and mpv of the histogram are 2.11, 2.03, and 1.52, respectively. For $\kappa=1$ (self-similar evolution, where the CME minor radius $a \propto r$), they are 1.72, 1.59, and 1.35, respectively. For $\kappa=1.5$, they are 1.56, 1.52, and 1.28, respectively. For $\kappa=2$, they are 1.45, 1.41, and 1.23, respectively.} 
\label{Fig: delta_nonselfsimilar}
\end{figure}

If the CME did not expand in a self-similar manner; i.e., if the minor radius ($a$) were to follow $a \propto r^{\kappa}$ (instead of $a \propto r$ as we have assumed, $r$ being the radial distance of the CME from the Sun), the change in CME volume, $dV/dr$, would be $= (2 \kappa + 1) V/r$ instead of $3V/r$ and the number 3 in the second term on the right hand side of Eq~\ref{nunlikeVasquez1} would be replaced with $2 \kappa + 1$. The histograms in Fig~\ref{Fig: delta_nonselfsimilar} show how $\langle \delta \rangle_{MO}$ varies with changes in $\kappa$. We adopt four different values for $\kappa$ ($\kappa = 0.5, 1.0, 1.5,$ and $2.0$). We note that $\kappa = 1$ corresponds to self-similar expansion (i.e., $a \propto r$). For $\kappa = 2$, the mpv of $\langle \delta \rangle_{MO}$ is 8 \% lower than that for $\kappa = 1$. For $\kappa = 1.5$, the mpv of $\langle \delta \rangle_{MO}$ is 5 \% lower than that for $\kappa = 1$, while it is 13\% higher for $\kappa = 0.5$. Our findings therefore suggest that the polytropic index decreases with the cross-sectional expansion of the CMEs (see Fig~\ref{Fig: delta_nonselfsimilar}). 
If the volume evolution were more involved than a situation which can be accomodated by the $a \propto r^{\kappa}$ prescription, we would have to prescribe $dV/dr$ accordingly in going from Eq~\ref{firstmod} to Eq~\ref{firstmod2}. 

Eq~\ref{reqdelta} also shows that $\langle \delta \rangle_{MO}$ is approximately proportional to $\langle \epsilon_{t} \rangle$. This suggests that the closer $\langle \delta \rangle_{MO}$ is to its isothermal value of 1, the lower the requirement for local heating via turbulent dissipation. This is expected because, if the protons in the MO are close to isothermal, it implies that they are thermally well connected to an external heat reservoir (such as the solar corona). Good thermal ``connection'' would imply a high thermal conductivity. Careful estimates of thermal conductivity in the turbulent, collisionless solar wind plasma would be needed to justify this. 

\begin{figure}
	
	\includegraphics[width=0.5\textwidth , scale=0.9]{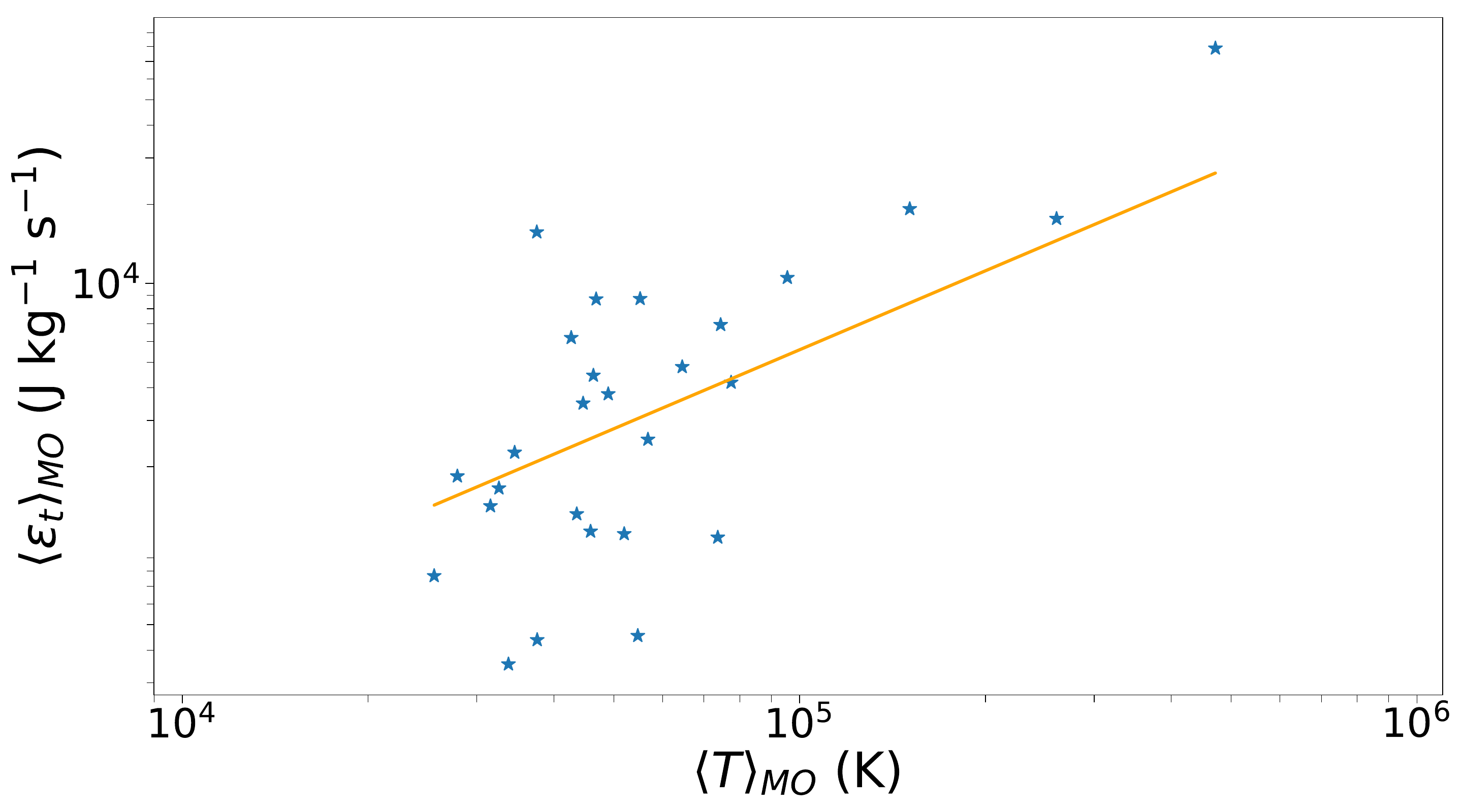}
	
	\caption{A scatterplot between the mean temperature ($\langle T \rangle_{\rm MO}$) and average turbulent energy cascade rate ( $\langle \epsilon_{\rm t} \rangle_{\rm MO}$) inside the MOs associated with the events listed in Table~\ref{T:Data}. The equation of the straight line fit is $y = 0.056x - 0.125$. The Pearson's correlation coefficient ($r$) is 0.93 with a p-value of $2.04 \times 10^{-12}$.} 
\label{Fig: epstempcorr}
\end{figure}

\section{Summary and Conclusions}
\label{S:Discussion}
\subsection{Summary}
CMEs are observed to expand as they propagate through the heliosphere. Such expansion will be accompanied by plasma cooling, with heating/cooling processes being typically treated (implicitly) via a polytropic index ($\delta$). If the expansion is adiabatic ($\delta$ close to 5/3), the CME plasma temperature will be as low as a few tens of Kelvin when it reaches the Earth. Since observed proton temperatures in near-Earth CMEs are $\approx 10^{5}$ K, this calls for significant local heating as CMEs propagate. If, on the other hand, the CME plasma maintains significant thermal contact with the solar corona, it can be expected to be nearly isothermal ($\delta \approx 1$) and the local heating requirement is minimal. Since the CME magnetic fields are usually considered to be the basic energy reservoir, a higher demand on local heating translates to a lesser energy available for propagation, and vice versa. The value of $\delta$ adopted in a CME propagation model thus substantially affects the predictions of the Sun-Earth propagation time and the Earth arrival speed. 
In this paper, we obtain an expression relating the local heating rate ($\epsilon$) corresponding to the polytropic index $\delta$ (Eq~\ref{new1}). The quantity $\epsilon$ can be regarded as the local heating rate required by a model that uses a given polytropic index $\delta$. On the other hand, the turbulent fluctuations in CMEs are often invoked as a plausible source of proton heating. Using observations of 27 well-observed geoeffective CMEs in solar cycle 24, we compute the power ($\epsilon_{t}$) available in the turbulent velocity fluctuations (Eq~\ref{turbpower}). The quantity $\epsilon_{t}$ can be thought of as the power that is potentially available to heat the CME plasma. Our main findings upon comparing $\epsilon$ with $\epsilon_{t}$ are as follows:

\begin{itemize}
\item
The available power ($\epsilon_{t}$) is $\approx 54 \%$ lower than the power ($\epsilon$) required with $\delta = 5/3$ (Fig~\ref{figead}).
\item
If, instead, the required power ($\epsilon$) is taken to be equal to the available power ($\epsilon_{t}$), the mpv for $\delta$ is 1.35 (Fig~\ref{Fig: hist}).

Interestingly, we find a strong correlation between the average dissipation rate ($\langle \epsilon_{t} \rangle$) and the temperature ($\langle T \rangle$) in the CME (Fig~\ref{Fig: epstempcorr}). The Pearson correlation coefficient between $\langle \epsilon_{t} \rangle$ and $\langle T \rangle$ is 0.93, with a low p-value of $10^{-12}$, which implies that the correlation is significant and reliable. We note that our results regarding the comparison of $\epsilon_{t}$ with $\epsilon$ and the estimate of $\langle \delta_{MO}\rangle$ are valid only near the Earth. In principle, Eq~\ref{nunlikeVasquez1} holds for any heliocentric distance $R$, provided the CME velocity $U$ and temperature $T$ is known. 
If the correlation between $\langle \epsilon_{t} \rangle$ and $\langle T \rangle$ generally holds true, it will allow us to estimate $\delta_{MO}$ for heliocentric distances between the Sun and the Earth, using reasonable models for $T(R)$ and $U(R)$.
\end{itemize}

\subsection{Discussion}

Estimates of CME arrival time and velocity (at the Earth) are vital inputs to space weather prediction. CME propagation models that calculate these quantities usually invoke CME magnetic fields and turbulent fluctuations as energy reservoirs. The polytropic index is often taken to be an adjustable parameter in such models. Our results show the relation between the polytropic index ($\delta$), CME velocity ($U$), proton temperature ($T$), and the local energy dissipation rate ($\epsilon_{t}$).



Assuming that the power in the turbulent velocity fluctuations partially provides for the work expended in CME expansion close to the Earth, we arrive at an estimate for the mpv of the polytropic index in near-Earth CMEs, $\langle \delta \rangle_{\rm MO} = 1.35$ . This can be taken to be a suggestion for the polytropic index to be adopted in CME propagation models. The most probable estimate of $1.35$ is likely to be an upper limit for $\delta$, for not all the power in the turbulent velocity fluctuation spectrum is likely to be expended in proton heating.
Incidentally, \cite{1996KumarRust} obtain $\delta = 1.33$ by assuming that the energy input ($dQ$) to the CME is a fraction of the energy contained in the CME magnetic fields (without specifying the mechanism by which magnetic energy is converted into plasma heating).

We have so far have used assumed values of $T_{0} = 10^{6}$ K at $R = 1.05 R_{\odot}$ and observed values for the CME velocity ($U$), velocity fluctuations ($\Delta U_{k}$) and proton temperature ($T$) to calculate $\alpha$ (Eq~\ref{powerlaw3}), $\langle \epsilon_{t} \rangle$ (Eq~\ref{turbpower}) and $\delta$ (Eq~\ref{reqdelta}). However, it is also worth examining some of the parametric dependencies from Eq~\ref{reqdelta}. These are:
\begin{itemize}
    \item 
    All else remaining fixed, the CME velocity increases with a decrease in $\delta$. Specifically, using fiducial values of $T_{0} = 10^{6}$ K, $R = 1.05 R_{\odot}$, $\langle \epsilon_{t} \rangle = 10^{4} {\rm J \, kg^{-1}\,s^{-1}}$, Eq~\ref{reqdelta} shows that decreasing $\delta$ from 1.38 to 1.3 (a decrease of 5\%) results in an increase in the CME velocity (at $R = 215 R_{\odot}$) from 250 ${\rm km\,s^{-1}}$ to 500 ${\rm km\,s^{-1}}$ (an increase of 100\%). This confirms what CME modelers recognize; assuming a thermodynamic equation of state that is close(r) to isothermal generally results in faster CMEs. These findings are also in agreement with those of \cite{2024khuntia2}.
    \item 
    All else remaining fixed, a decrease in $\delta$ results in a decrease in $\alpha$ (Eq~\ref{powerlaw3}), which means that the proton temperature decreases slower with heliocentric distance. Using fiducial values of $T_{0} = 10^{6}$ K, $R = 1.05 R_{\odot}$, $\langle \epsilon_{t} \rangle = 10^{4} {\rm J \, kg^{-1}\,s^{-1}}$, and a CME velocity of 400 ${\rm km\,s^{-1}}$ at $R = 215 R_{\odot}$, Eq~\ref{reqdelta} shows that decreasing $\delta$ from 1.38 to 1.3 (a decrease of 5\%) results in $\alpha$ decreasing from 0.85 to 0.6 (a decrease of 29\%). In other words, the rate of decrease of the proton temperature (with heliocentric distance) is slower. This will be manifested as a relatively higher proton temperature at the Earth. This trend is borne out by solar wind simulations (e.g., \cite{2022Prateek}) where the predicted proton temperature at the Earth is higher for polytropic indices that are closer to 1. 
\item 

As mentioned in the discussion following Eq~\ref{turbpower} in \S~\ref{S-power in turbulence}, including the power potentially available in magnetic fluctuations (in addition to the power potentially available in velocity fluctuations) results in the quantity $C_{0}$ being replaced by $2^{3/2} C_{0}$. Using $\langle \epsilon_{t} \rangle = 2^{3/2} \times 10^{4} {\rm J \, kg^{-1}\,s^{-1}}$, and a CME velocity of 400 ${\rm km\,s^{-1}}$ at $R = 215 R_{\odot}$ in Eq~\ref{reqdelta} gives $\delta = 1.45$ (which can be regarded as an increase of $\approx 7.4$\% over the mpv of $\delta = 1.35$ which was obtained without the factor of $2^{3/2}$). 
    
\item 

An alternative approach to the one we have taken in this paper would be to observationally determine the polytropic index in MOs and their sheaths using fits to a $T \propto n^{\delta - 1}$ scatterplot (e.g., \cite{2025katsavrias}). Eq~\ref{reqdelta} could then be used to compute the power that needs to be dissipated on the protons, as implied by this value of the polytropic index.
\end{itemize}

\section*{Acknowledgements}
The authors thank the Indo-US Science and Technology Forum (IUSSTF) for supporting the current project. DB acknowledges the support from the University of Glasgow to meet the Open Access Publication of this paper. This paper has benefited from helpful inputs from the anonymous reviewer.




\section*{Data Availability}

The main data used in this article are available on the WIND website (\url{ht
tps://wind.nasa.gov/}). The data detailing the event list and estimated
parameters are available in this article and can be used with proper
citations.



\bibliographystyle{mnras}
\bibliography{example} 




\appendix
\label{appendix}
\section{Appendix}
\subsection{Relating the proton polytropic index to the local heating rate}
We modify the treatment of \cite{1995Verma} and \cite{2007Vasquez} to obtain an equation for the evolution of proton temperature in a CME by accounting for a polytropic equation of state. As in \cite{1995Verma}, we start with the energy conservation equation
\begin{equation}
dQ = dU^{'} + P dV \, ,
\label{firstlaw}
\end{equation}
where $dQ$ is the change in energy added to (or taken from) the CME, $dU^{'}$ is the change in internal energy, $P dV$ is the work done in expanding/contracting the CME volume. The change in internal energy can be written as 
\begin{equation}
dU^{'} = N C_{v} dT
\label{eqdU}
\end{equation}
where $V$ and $N$ are the volume of the CME and the number of moles contained in it, $dT$ is the change in temperature and $C_{v}$ is the specific heat at constant volume. Introducing $K \equiv dQ/P dV$ and using Eq~\ref{eqdU}, Eq~\ref{firstlaw} can be rewritten as
\begin{equation}
(K - 1) P dV = dU^{'} = N C_{v} dT \, .
\label{firstlawrewrite}
\end{equation}
We introduce the polytropic law 
\begin{equation}
P V^{\delta} = C \, 
\label{polytropic}
\end{equation}
and differentiate it to get
\begin{equation}
\frac{dP}{P} + \delta \frac{dV}{V} = 0 \, .
\label{polydiff}
\end{equation}
Similarly, differentiating the the ideal gas law $P V = N R T$ (where $R \equiv C_{p} - C_{v}$) and using Eq~\ref{firstlawrewrite} gives

\begin{equation}
[ 1 - (\gamma - 1) (K - 1) ] \frac{dV}{V} + \frac{dP}{P} = 0 \, ,
\label{eq3}
\end{equation}
where $\gamma \equiv C_{p}/C_{v}$ is the usual adiabatic index. Comparing Eq~\ref{eq3} and Eq~\ref{polydiff} gives

\begin{equation}
K \equiv \frac{dQ}{p dV} = \frac{\delta - \gamma}{1 - \gamma} \, .
\label{Kdef}
\end{equation}
The energy conservation equation (Eq~\ref{firstlaw}) does not include ``external'' contributions from sources such as turbulent dissipation. Accounting for such contributions and using Eqs~\ref{eqdU} and \ref{Kdef}, the energy conservation equation becomes
\begin{equation}
M \epsilon dt = d U^{'} + P dV - d Q = N C_{v} dT + P dV \frac{1 - \delta}{1 - \gamma}
\label{firstmod}
\end{equation}
The quantity $M$ represents the CME mass, $dt$ the time elapsed and $\epsilon$ (${\rm erg\,g^{-1}\,s^{-1}}$) the time rate per unit mass of additional energy deposition. The additional energy deposition term $M\,\epsilon\, dt$ is not accounted for in the original energy conservation equation (Eq~\ref{firstlaw}). It is ``additional'' in the sense that it represents heating due to additional sources such as small scale reconnection events inside the CME plasma, or from dissipation of turbulent fluctuations.
We next differentiate Eq~\ref{firstmod} with respect to $r$ and write $dr/dt \equiv U(r)$, so that $U(r)$ represents the CME velocity. If the CME flux rope is idealized as a curved cylinder of cross-sectional radius $a$ and length $\propto r$, it's volume $V \propto a^{2}r$. Since most CMEs expand in a self-similar manner with $a \propto r$, it follows that $V \propto r^{3}$, which means that $dV/dr = 3 V/r$. This yields
\begin{equation}
\frac{M}{N C_{v}}\frac{\epsilon}{U} = \frac{dT}{dr} + 3 \frac{1 - \delta}{1 - \gamma} \frac{P}{N C_{v}} \frac{V}{r} \, .
\label{firstmod2}
\end{equation}
Using the ideal gas law ($PV = NRT$), $R \equiv C_{p} - C_{v}$ and $M/(N C_{v}) = (\gamma - 1) m_{p}/k_{B}$ ($m_{p}$ is the proton mass and $k_{B}$ is the Boltzmann constant) in Eq~\ref{firstmod2}, we get
\begin{equation}
\frac{dT(R)}{dR} = 84.32 \, (\gamma - 1)\frac{\epsilon(R)}{U(R)} - 3 (\delta - 1) \frac{T(R)}{R} \, ,
\label{nunlikeVasquez}
\end{equation}
where $R = r/R_{\odot}$ and $U(R)$ is the CME velocity in units of ${\rm km \, s^{-1}}$. If we use $\delta = \gamma = 5/3$ and 2 in place of 3 in the second term on the right hand side, Eq~\ref{nunlikeVasquez} is identical to Eq 9 of \cite{2007Vasquez}. \\

\subsection{Relevant Tables}
This section contains two tables. Table~\ref{T:Data} gives details of the 27 events that are used in this paper. Table~\ref{T:EpsilonDelta} lists $\epsilon$ (Eq~\ref{new1}), $\langle\epsilon_{t}\rangle_{MO}$ (Eq~\ref{turbpower}), $\langle\delta\rangle_{MO}$ (Eq~\ref{reqdelta}), and $E_{\rm ad}$ (Eq~\ref{Ead}) for all the events listed in Table~\ref{T:Data}.


\begin{landscape}

\begin{table}
	
	\caption{
		The list of the near-Earth ICME events we use in this study. The arrival date and time of the ICME at the position of the WIND measurement and the arrival and departure dates \& times of the associated magnetic clouds (MOs) are taken from the WIND ICME catalogue (\url{wind.nasa.gov/ICMEindex.php}). The time between the `ICME start time' and `MO start time' is taken as the sheath duration.}  
	\begin{center}
		
		\begin{tabular}{cccclccc}
			
			\hline
			ICME        & ICME start date & MO start  & MO end       & Flux  & $D_{\rm st}$\\
			event      & and time[UT]   &  date and   & date and     & rope & index \\
			number    &  (1AU)         &  time [UT]   &   time [UT]   &   type   & (nT)\\
			\hline
			\hline
			
			1    &  2010 04 05 , 07:55 & 2010 04 05 , 11:59 & 2010 04 06 , 16:48 & Fr  & -81 \\
			2    &  2010 05 28 , 01:55 & 2010 05 28 , 19:12 & 2010 05 29 , 17:58 & Fr  & -80\\
			3    &  2011 05 28 , 00:14 & 2011 05 28 , 05:31 & 2011 05 28 , 22:47 & F+  & -80\\
			4    &  2011 10 24 , 17:41 & 2011 10 25 , 00:21 & 2011 10 25 , 23:31 & Cx  & -147\\
			5    &  2012 03 08 , 10:32 & 2012 03 08 , 19:55 & 2012 03 11 , 07:26 & Cx & -145\\
			6    &  2012 06 16 , 09:03 & 2012 06 16 , 22:01 & 2012 06 17 , 11:23 & F+ &  -86\\
			7    &  2012 07 08 , 02:10 & 2012 07 08 , 07:58 & 2012 07 10 , 01:41 & Cx  & -78 \\
			8    &  2012 07 14 , 17:39 & 2012 07 15 , 06:14 & 2012 07 17 , 03:21 & Fr  & -139 \\ 
			9    &  2012 09 30 , 10:14 & 2012 09 30 , 12:14 & 2012 10 02 , 02:53 & Cx & -122\\
			10   &  2012 10 08 , 04:12 & 2012 10 08 , 15:50 & 2012 10 09 , 17:17 & Fr  & -107\\

			11   &  2012 10 12 , 08:09 & 2012 10 12 , 18:29 & 2012 10 13 , 09:14 & Fr  & -90\\
			
			12   &  2012 11 12 , 22:12 & 2012 11 13 , 08:23 & 2012 11 14 , 08:09 & F+ & -108\\

			13   &  2013 03 17 , 05:21 & 2013 03 17 , 14:09 & 2013 03 19 , 16:04 & Fr & -132\\
			
			14   &  2013 03 06 , 02:09 & 2013 06 06 , 14:23 & 2013 06 08 , 00:00 & F+ & -78\\
			
			15   &  2013 06 27 , 13:51 & 2013 06 28 , 02:23 & 2013 06 29 , 11:59 & Fr & -102\\ 
			16   &  2013 07 04 , 17:17 & 2013 07 05 , 04:05 & 2013 07 07 , 14:24 & Cx & -87\\
			17   &  2014 09 12 , 15:17 & 2014 09 12 , 21:22 & 2014 09 14 , 11:38 & F- & -88\\
			
			18   &  2015 01 07 , 05:38 & 2015 01 07 , 06:28 & 2015 01 07 , 21:07 & F+ & -99\\
			
			19   &  2015 06 22 , 18:07 & 2015 06 23 , 02:23 & 2015 06 24 , 13:03 & Cx & -198\\
			
			20   &  2015 11 06 , 17:46 & 2015 11 07 , 07:11 & 2015 11 08 , 16:47 & Fr & -79\\
			
			21   &  2015 12 19 , 15:35 & 2015 12 20 , 13:40 & 2015 12 21 , 23:02 & Fr & -155\\

			22   &  2016 01 19 , 03:31 & 2016 01 19 , 11:23 & 2016 01 20 , 14:19 & Fr & -93\\
			23   &  2016 10 12 , 21:37 & 2016 10 13 , 06:27 & 2016 10 14 , 16:19 & F+ & -110\\
			24   &  2017 05 27 , 13:45 & 2017 05 27 , 22:50 & 2017 05 29 , 11:05 & F+ & -125\\
			25   &  2017 09 06 , 22:21 & 2017 09 07 , 09:13 & 2017 09 08 , 06:35 & Ej & -122\\
			26   &  2017 09 07 , 16:17 & 2017 09 08 , 00:19 & 2017 09 09 , 16:57 & Ej & -109\\
			27   &  2018 08 25 , 01:02 & 2018 08 25 , 12:04 & 2018 08 25 , 12:19 & F+ & -175\\
			
			\hline
			
		\end{tabular}
	\end{center}
	\label{T:Data}
\end{table}
\end{landscape}

\begin{landscape}
\begin{table}

	\caption{The first column is the serial number of the event, taken from Table~\ref{T:Data}. The second column is the average proton temperature inside the MO. The third column is the required power ($\epsilon$) using Eq~\ref{new1}. The fourth column is the average turbulent cascade rate ($\langle\epsilon_{t}\rangle_{MO}$). The fifth column is the average polytropic index ($\langle \delta \rangle_{MO}$) obtained from Eq~\ref{reqdelta}. The sixth column is the percentage difference between $\epsilon$ and $\langle\epsilon_{t}\rangle_{MO}$ using Eq~\ref{Ead}. }
	
	\centering
	\begin{tabular}{cccccc}
			\hline
	ICME        & Average & $\epsilon$  & $\langle\epsilon_{t}\rangle_{MO}$       & $\langle\delta \rangle_{MO}$  & $E_{\rm ad}$ (\%) \\
	event      & proton temperature in MO  &   & $t_{\rm box} = 60$ min     & $t_{\rm box} = 60$ min & $t_{\rm box} = 60$ min \\
	number    &  (K)       & (J kg$^{-1}$ s$^{-1}$)   & (J kg$^{-1}$ s$^{-1}$)   & (at 1 AU)  & (at 1 AU) \\
	\hline
	\hline
	1 & 42661.07 & 3154.98 & 6198.88 & 2.12 & 96.48 \\ 
	2 & 37577.24 & 1522.35 & 437.40 & 1.35 & -71.27 \\ 
	3 & 44599.27 & 2712.16 & 3486.56 & 1.80 & 28.55 \\ 
	4 & 95528.45 & 6100.42 & 10501.63 & 2.06 & 72.15 \\ 
	5 & 77455.27 & 5599.40 & 4188.05 & 1.54 & -25.21 \\ 
	6 & 64545.57 & 3562.04 & 4802.49 & 1.84 & 34.82 \\ 
	7 & 32580.03 & 1523.78 & 1656.03 & 1.71 & 8.68 \\ 
	8 & 48978.26 & 2852.42 & 3614.58 & 1.59 & 32.66 \\ 
	9 & 56799.44 & 2427.59 & 2540.65 & 1.61 & -19.13 \\ 
	10 & 46820.43 & 2252.08 & 8700.73 & 3.01 & 286.23 \\ 
	11 & 55167.73 & 3189.94 & 8729.32 & 2.51 & 173.65 \\ 
	12 & 73696.42 & 3509.89 & 1075.31 & 1.35 & -69.63 \\ 
	13 & 46340.17 & 2885.91 & 4453.36 & 1.92 & 54.32 \\ 
	14 & 33756.76 & 1629.67 & 353.49 & 1.35 & -78.39 \\ 
	15 & 25589.45 & 1086.06 & 767.25 & 1.54 & -29.38 \\ 
	16 & 27901.36 & 1040.41 & 1840.02 & 1.89 & 76.86 \\ 
	17 & 74488.79 & 5546.86 & 6947.34 & 1.76 & 25.25 \\ 
	18 & 31567.11 & 1589.26 & 1246.12 & 1.57 & -19.12 \\ 
	19 & 150787.24 & 12288.69 & 9877.82 & 1.56 & 56.32 \\ 
	20 & 43552.77 & 2692.16 & 1320.31 & 1.43 & -50.96 \\ 
	21 & 45858.42 & 2148.55 & 1133.49 & 1.42 & -47.24 \\ 
	22 & 37522.46 & 1556.93 & 4473.13 & 2.53 & 105.96 \\ 
	23 & 34547.13 & 1544.09 & 2265.36 & 1.80 & 46.72 \\ 
	24 & 54687.12 & 2360.98 & 453.71 & 1.28 & -80.78 \\ 
	25 & 471530.44 & 42419.50 & 21387.22 & 1.35 & 85.28 \\ 
	26 & 260728.26 & 25678.55 & 8641.99 & 1.28 & -31.36 \\ 
	27 & 51986.37 & 2527.93 & 1108.91 & 1.35 & -46.14 \\
	\hline
	
	\end{tabular}
	\label{T:EpsilonDelta}
\end{table}
\end{landscape}


\bsp	
\label{lastpage}
\end{document}